\documentclass[
 reprint,
amsmath,amssymb,
pra,
]{revtex4-1}

\usepackage[T1]{fontenc}

\usepackage{graphicx}

\usepackage{dcolumn}
\usepackage{bm}
\usepackage{hyperref}
\usepackage{color}

\begin{document}

\preprint{APS/123-QED}

\title{Normal-mode splitting in the optomechanical system with an optical parametric 
amplifier and coherent feedback}
\author{Yue Li}
\affiliation{
State Key Laboratory of Quantum Optics and Quantum Optics Devices, Institute of Opto-Electronics, Shanxi University, Taiyuan 030006, China
}
\author{Hengxin Sun}%
 \email{hxsun@sxu.edu.cn}
\affiliation{
State Key Laboratory of Quantum Optics and Quantum Optics Devices, Institute of Opto-Electronics, Shanxi University, Taiyuan 030006, China
}
\affiliation{
Collaborative Innovation Center of Extreme Optics, Shanxi University, Taiyuan 030006, China
}
\author{Yijian Wang}%
\affiliation{
State Key Laboratory of Quantum Optics and Quantum Optics Devices, Institute of Opto-Electronics, Shanxi University, Taiyuan 030006, China
}
\author{Kui Liu}%
\affiliation{
State Key Laboratory of Quantum Optics and Quantum Optics Devices, Institute of Opto-Electronics, Shanxi University, Taiyuan 030006, China
}
\affiliation{
Collaborative Innovation Center of Extreme Optics, Shanxi University, Taiyuan 030006, China
}
\author{Jiangrui Gao}%
\affiliation{
State Key Laboratory of Quantum Optics and Quantum Optics Devices, Institute of Opto-Electronics, Shanxi University, Taiyuan 030006, China
}
\affiliation{
Collaborative Innovation Center of Extreme Optics, Shanxi University, Taiyuan 030006, China
}
\date{\today}
\begin{abstract}
Strong coupling in optomechanical systems is the basic condition for observing many quantum phenomena such as optomechanical squeezing and entanglement. Normal-mode splitting (NMS) is the most evident signature of strong coupling systems. Here we show the NMS in the spectra of the movable mirror and the output field in an optomechanical system can be flexibly engineered by a combination of optical parametric amplifier (OPA) and coherent feedback (CF). Moreover, the NMS could be enhanced by optimizing the parameters such as input optical power, OPA gain and phase, CF strength in terms of amplitude reflectivity of beam splitter.
\end{abstract}

\maketitle

\section{\label{introduction}Introduction}

In recent decades, cavity optomechanics composed of coupled cavity field and movable mirror, has become an important field due to its potential applications in quantum optics \cite{Schwab2005,LaHaye2004,Loudon1981,Gigan2006}. A prerequisite for these applications is ground state cooling of the movable mirror. Recently, great progress has been made in achieving ground state cooling of oscillators with various methods, such as dispersive coupling \cite{Teufel2011,Arcizet2006,Groeblacher2009,Rocheleau2010}, dissipative coupling \cite{Elste2009,Xuereb2011}, dynamic cooling \cite{Li2011,Liu2013}, atom-assisted cooling \cite{Genes2009,Restrepo2014,Vogell2013}, and external cavity cooling \cite{Xuereb2010}, which sets the stage for us to observe the quantum behavior such as mechanical squeezing \cite{Mari2009,Liao2011,Wollman2015}, mechanical entanglement \cite{Ockeloen-Korppi2018}, and optomechancal squeezing \cite{Mancini1994,Fabre1994,Marino2010,Purdy2013a,Aggarwal2020} and entanglement \cite{Purdy2017,Sudhir2017a,Yu2020,Chen2020}.

Normal-mode splitting (NMS) is the most evident signature in strong coupling optomechanical systems \cite{Marquardt2007,Dobrindt2008,Huang2009,Bhattacherjee2009,Kumar2010,Yong-Chun2015,Han2013,Rossi2018,Huang2019,Zhang2019}. The NMS generally occurs when the energy exchanging rate between two coupled subsystems is much faster than their energy-dissipating into the environment. The concept of NMS originally comes from the vacuum Rabi splitting in a coupled atom-cavity system in 1980s \cite{Sanchez-Mondragon1983,Agarwal1984,Thompson1992,Reithmaier2004}. The NMS exhibits two-peak spectra of the position of movable mirror and the noise spectra of output optical field in cavity optomechanical systems \cite{Marquardt2007,Dobrindt2008}, basically due to strong coupling. There are several methods to enhance the NMS effect. Particularly, enhancement could be realized by adding a degenerate optical parametric amplifier (OPA) \cite{Huang2009} in the cavity, or introducing a single coherent feedback (CF) \cite{Huang2019} outside the cavity.  Here, we combine the two schemes of OPA and CF and analyze the NMS. Compared to the previous scheme with OPA or CF alone, more flexible degrees of freedom could be utilized to control the optomechanical coupling strength and NMS. Strong coupling and more obvious NMS could be achieved by optimizing the parameters such as input optical power, OPA gain and phase, and amplitude reflectivity of the beam splitter of CF.

The layout of the paper is presented below. In Sec. \ref{model} we introduce the theoretical model, present the Hamiltonian of the system, give the Langevin equations of motion for the movable mirror and the cavity field, and obtain the steady-state mean values. In Sec. \ref{radiation pressure} we linearize the quantum Langevin equations, give the stability conditions of the system, derive the spectrum of position fluctuation of movable mirror. In Sec. \ref{normal mode splitting} we analyze the behavior of the NMS in terms of location and linewidth of two normal modes by varying the following parameters: amplitude reflectivity of beam splitter, input laser power, OPA gain and phase, and compare it with the case only OPA or CF is added. In Sec. \ref{spectrum} we get the spectrum of output field and show the two-peak spectra of movable mirror and output field.

\section{Model}\label{model}
\begin{figure}[t]
\includegraphics[width=7cm]{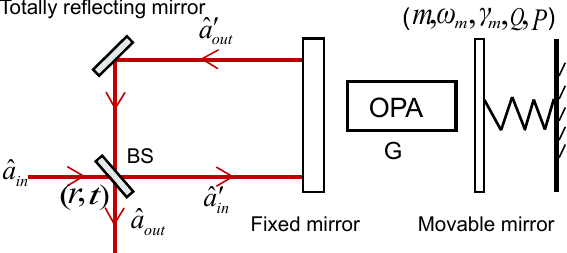}
\caption{Opto-mechanical system with an OPA and CF. The transmitted part of the input laser is sent into the cavity by a fixed mirror. Then a part of output field from the cavity field is fed back into the cavity through a totally reflecting mirror and a partially reflecting beam splitter (BS).}
\label{FIG.1l}
\end{figure}

As shown in Fig. \ref{FIG.1l}, we consider an optical cavity consisting of two mirrors separated by a distance $L$, composed of one fixed mirror with partial power reflectivity and one movable mirror with total power reflectivity. A second-order nonlinear OPA device is placed in the cavity. The cavity output field from the fixed mirror is partially sent back into the cavity via a totally reflecting mirror and a beam splitter (BS), forming an optical coherent feedback. The movable mirror is in a thermal bath at temperature $T$ and regarded as a quantum mechanical harmonic oscillator with effective mass $m$, resonance frequency $\omega_m$, and damping rate $\gamma_m$. An input laser beam with frequency $\omega_l$ and an amplitude $\varepsilon_l$ related to a power of ${{P}_{in}}$ by ${{\varepsilon }_{l}}=\sqrt{\frac{2\kappa {{P}_{in}}}{\hbar {{\omega }_{l}}}}$,  is split into two parts by the BS with amplitude reflectivity $r$ and transmissivity $t$, $\kappa$ is the cavity field decay rate from the fixed mirror and $\hbar$ is Planck constant divided by $2\pi$. No extra optical loss is assumed. The cavity field exerts a radiation pressure force on the movable mirror due to momentum transfer from the photons in the cavity. The position of the movable mirror oscillates around its equilibrium position under the thermal Langevin force and the radiation pressure force. 

The adiabatic limit, $\omega _{m}\ll \mathit{}\pi{c} /L$  is assumed, where $c$  is the light speed in vacuum and $L$ is the cavity length. Hence, we can consider the model to the case of single-cavity and mechanical mode \cite{Law1994,Law1995}. In the frame rotating at the laser frequency $\omega_l$, the total Hamiltonian describing the coupled system is given by
 \begin{align}\label{eq:1}
   \hat{H}=&\hbar ({{\omega }_{c}}-{{\omega }_{l}}){{{\hat{a}}}^{\dagger }}\hat{a}-\hbar {{g}_{0}}{{{\hat{a}}}^{\dagger }}\hat{a}\hat{Q}+\frac{\hbar {{\omega }_{m}}}{4}({{{\hat{Q}}}^{2}}+{{{\hat{P}}}^{2}})\nonumber\\
 & +i\hbar t{{\varepsilon }_{l}}({{{\hat{a}}}^{\dagger }}-\hat{a})+i\hbar G({{e}^{i\theta }}{{{\hat{a}}}^{\dagger 2}}-{{e}^{-i\theta }}{{{\hat{a}}}^{2}}),
 \end{align}
where $\hat a$ (${\hat a^\dag }$) is the annihilation (creation) operator of the fundamental cavity field, $Q$ and $P$ are the dimensionless position and momentum operators of the movable mirror with $\hat{Q}=\sqrt{\frac{2 m \omega_{m}}{\hbar}} \hat{q}$, $\hat{P}=\sqrt{\frac{2}{m \hbar \omega_{m}}} \hat{p}$, and they obey the relationship ${[\hat{Q}, \hat{P}]=2 i }$. In Eq. \eqref{eq:1}, the first term represents the energy of the cavity field, ${\hat{n}_{a}}={{\hat{a}}^{\dagger }}\hat{a}$ is the number of the photons inside the cavity. The second term describes the optomechanical interaction between cavity field and movable mirror via radiation pressure, ${{g}_{0}}=\frac{{{\omega}_{c}}}{L}\sqrt{\frac{\hbar }{2m{{\omega }_{m}}}}$ is the single-photon optomechanical coupling constant. The third term describes the energy of movable mirror. The fourth term corresponds to the cavity field driven by the external field. The last term denotes the second-order nonlinear interaction energy, $G$ is the OPA gain related to the power of second harmonic field, $\theta$ is the relative phase between the fundamental and second harmonic fields. 

Using the Heisenberg equations of motion, adding the noise and  damping terms, and also taking the feedback term into account, we obtain the following Langevin equations of motion:
\begin{subequations}\label{eq:group1}
\begin{gather}
\dot{\hat{Q}}={{\omega }_{m}}\hat{P}, \label{eq:a} \\
\dot{\hat{P}}=2{{g}_{0}}{\hat{n}_{a}}-{{\omega }_{m}}\hat{Q}-{{\gamma }_{m}}\hat{P}\text{+}\hat{\xi }, \label{eq:b} \\
\dot{\hat{a}}=-i( {{\omega }_{c}}-{{\omega }_{l}}-{{g}_{0}}\hat{Q} )\hat{a}+2G{{e}^{i\theta }}{{{\hat{a}}}^{\dagger }}-\kappa \hat{a} \nonumber\\ 
  +t{{\varepsilon }_{l}}+\sqrt{2\kappa }( t\delta {{{\hat{a}}}_{in}}+r{{{\hat{{a}'}}}_{out}} ). \label{eq:c} 
\end{gather}
\end{subequations}
where the first two terms are related to the position and momentum of the movable mirror, respectively, while the third one corresponds to the intracavity field.\\
The force $\hat{\xi }$ is related to the thermal noise of the movable mirror in thermal equilibrium, which has zero mean value and nonzero time domain correlation function \cite{Giovannetti2001}
\begin{align}\label{eq:3}
   \left\langle \xi (t)\xi ({t}') \right\rangle =&\frac{\hbar {{\gamma }_{m}}}{2\pi }m\int{d\omega \omega {{e}^{-i\omega (t-{t}')}}} \nonumber\\ 
 & \times\Bigl[\coth \left( \frac{\hbar {{\omega }_{m}}}{2{{k}_{B}}T} \right)+1\Bigm],
\end{align}
where ${{k}_{B}}$ is the Boltzmann constant and $T$ is the environment temperature.\\ 
In Eq. \eqref{eq:c}, $\delta {{\hat{a}}_{in}}$ is the optical vacuum noise operator with zero mean value and its $\delta \text{-}$correlated function in the time domain \cite{Gardiner2004} is given by
\begin{subequations}\label{eq:group2}
\begin{gather}
\langle \delta{{{\hat{a}}}_{in}}(t)  \delta \hat{a}_{in}^{\dagger }({t}')\rangle =\delta (t-{t}'),\label{eq:4a}\\
\langle \delta {{{\hat{a}}}_{in}}(t) \delta {{{\hat{a}}}_{in}}({t}') \rangle= \langle \delta \hat{a}_{in}^{\dagger }(t) \delta {{{\hat{a}}}_{in}}({t}') \rangle =0.\label{eq:4b}
\end{gather}
\end{subequations}
According to the input-output relation of the cavity \cite{Collett1984}
\begin{equation}\label{eq:5}
{{\hat{{a}'}}_{out}}=\sqrt{2\kappa }\hat{a}-{{\hat{{a}'}}_{in}},
\end{equation}
and the beam splitter model
\begin{equation}\label{eq:6}
{{\hat{{a}'}}_{in}}=t{{\hat{a}}_{in}}+r{{\hat{{a}'}}_{out}},
\end{equation}
we get
 \begin{equation}\label{eq:7}
   {{\hat{{a}'}}_{out}}=\frac{\sqrt{2\kappa}}{1+r}\hat{a}-\frac{t}{1+r}{{\hat{a}}_{in}}.
 \end{equation}
We get the input-output relationship similar with Ref. \cite{Zhihui2011}, in which the CF is used to manipulate entanglement from a nondegenerate OPA. Note Eq. \eqref{eq:7} is different from the general input-output relation of the cavity as Eq. \eqref{eq:5}. Appendix \ref{appendix} gives detailed derivation of this relation by solving equations of field relations. Substituting Eq. \eqref{eq:7} into Eq. \eqref{eq:c}, we get the final motional equation of intracavity field:
\begin{align}\label{eq:8}
  \dot{\hat{a}}=&-[{{\kappa }_{eff}}+i({{\omega }_{c}}-{{\omega }_{l}}-{{g}_{0}}\hat{Q})]\hat{a}+t{{\varepsilon }_{l}}\Big (1-\frac{r}{r+1}\Bigl)\nonumber \\ 
 & \text{+}2G{{e}^{i\theta }}{{{\hat{a}}}^{\dagger }}+\sqrt{2\kappa }\Big(1-\frac{r}{r+1}\Bigl)t\delta {{{\hat{a}}}_{in}}.
\end{align}
Due to the CF, the effective cavity decay rate (or cavity linewidth) becomes ${{\kappa }_{eff}}=\kappa ({1-r})/({1+r})$, which decreases with increasing $r$ within a scale of $[-1,\ 1]$. Setting the time derivative terms in Eqs. \eqref{eq:a}, \eqref{eq:b} and \eqref{eq:8} to be zero, the steady-state solutions are given by

\begin{subequations}\label{eq:group3}
\begin{gather}
{{P}_{s}}=0, \label{eq:9a}\\
{{Q}_{s}}=\frac{2{{g}_{0}}}{{{\omega }_{m}}}{{\left| {{a}_{s}} \right|}^{2}},\label{eq:9b}\\
{{a}_{s}}=\frac{{{\kappa }_{eff}}-i\Delta +2G{{e}^{i\theta }}}{{{\kappa }_{eff}}^{2}+{{\Delta }^{2}}-4{{G}^{2}}}t{{\varepsilon }_{l}}\Big(1-\frac{r}{1+r}\Bigl)\label{eq:9c},
\end{gather}
\end{subequations}
where $\Delta ={{\omega }_{c}}-{{\omega }_{l}}-{{g}_{0}}{{Q}_{s}}$ is the effective cavity detuning where ${{g}_{0}}{{Q}_{s}}$ corresponds a constant frequency shift due to the radiation pressure of intra-cavity field exerts on the movable mirror. Note that the new equilibrium position ${{Q}_{s}}$ of the movable mirror depends on both optomechanical coupling constant ${g}_{0}$ and cavity photon number ${{\left| {{a}_{s}} \right|}^{2}}$. 

\section{Radiation pressure and quantum fluctuations}\label{radiation pressure}
In this section, we use the semi-classical method to obtain the quantum Langevin equations (QLEs) for quantum fluctuations and give the solutions for the quantum fluctuation and noise spectra of the position of movable mirror in the Fourier frequency domain. The operators are linearized as the sum of their average value and  fluctuation terms: $\hat{Q}={{Q}_{s}}+\delta \hat{Q}$, $\hat{P}={{P}_{s}}+\delta \hat{P}$, $\hat{a}={{a}_{s}}+\delta \hat{a}$, where $\delta \hat{P}$, $\delta \hat{Q}$, $\delta \hat{a}$ are respectively the small fluctuations of the position, momentum of the movable mirror and the cavity field. Then the QLEs are given by
\begin{subequations}\label{eq:group4} 
\begin{gather}
 \delta \dot{\hat{Q}}={{\omega }_{m}}\delta \hat{P},\label{eq:10a}\\
\delta \dot{\hat{P}}=2{{g}_{0}}(a_{s}^{*}\delta {{\hat{a}}^{\dagger }}+{{a}_{s}}\delta \hat{a})-{{\omega }_{m}}\delta \hat{Q}-{{\gamma }_{m}}\delta \hat{P}+\hat{\xi },\label{eq:10b}\\
\delta \dot{\hat{a}}=-({{\kappa }_{eff}}+i\Delta )\delta \hat{a}+i{{g}_{0}}\delta \hat{Q}{{a}_{s}}+2G{{e}^{i\theta }}\delta {{{\hat{a}}}^{\dagger }}\nonumber\\
+\sqrt{2\kappa }\Big(1-\frac{r}{r+1}\Bigl)t\delta {{{\hat{a}}}_{in}}.
\end{gather}
\end{subequations}

Generally, $g=2{{g}_{0}}{{a}_{s}}$ is referred to as “the light-enhanced optomechanical coupling strength”, which is proportional to the square root of input laser power. Introducing the amplitude and phase quadrature fluctuation operators of the cavity field and the input noise: $\delta x=\delta \hat{a}+\delta {{\hat{a}}^{\dagger }}$, $\delta y=i(\delta {{\hat{a}}^{\dagger }}-\delta \hat{a})$, $\delta {{x}_{in}}=\delta {{\hat{a}}_{in}}+\delta \hat{a}_{in}^{\dagger }$, $\delta {{y}_{in}}=i(\delta \hat{a}_{in}^{\dagger }-\delta {{\hat{a}}_{in}})$, we get the matrix form representation of Eq. \eqref{eq:group4}  as
\begin{equation}\label{eq:11}
\dot{\mu }(t)=M\mu (t)+\nu (t),  
\end{equation}
where $\mu (t)$ and $\nu (t) $ are the vectors of the fluctuation operators and input noise operators, respectively, 
\begin{equation}\label{eq:12}
\mu (t)={{\left( \delta \hat{P},\delta \hat{Q},\delta x,\delta y \right)}^{T}},
\end{equation}
\begin{equation}\label{eq:13}
\nu {{(t)}^{T}} =\bigg( 0,\hat{\xi },\sqrt{2\kappa }\Big(\frac{1}{1+r}\Bigl)\delta {{x}_{in}},\sqrt{2\kappa }\Big(\frac{1}{1+r}\Bigl)\delta {{y}_{in}} \biggl).
\end{equation}
The matrix $M$ is found to be
\begin{widetext}
\begin{equation}\label{eq:14}
M=\\\left( \begin{matrix}
   0 & {{\omega }_{m}} & 0 & 0  \\
   -{{\omega }_{m}} & -{{\gamma }_{m}} & {{g}_{0}}({{a}_{s}}+a_{s}^{*}) & -i{{g}_{0}}({{a}_{s}}-a_{s}^{*})  \\
   i{{g}_{0}}({{a}_{s}}-a_{s}^{*}) & 0 & 2G\cos \theta -{{\kappa }_{eff}} & 2G\sin \theta +\Delta   \\
   {{g}_{0}}({{a}_{s}}+a_{s}^{*}) & 0 & 2G\sin \theta -\Delta  & -(2G\cos \theta +{{\kappa }_{eff}})  \\
\end{matrix} \right).
\end{equation}
\end{widetext}
The stability of the system is determined by the eigenvalues of the matrix $M$. When all the eigenvalues of the matrix $M$ have negative real parts, the system is stable. According to the Routh-Hurwitz criterion \cite{Ed1964,Dejesus1987}, the stability conditions are given by
\begin{subequations}\label{eq:group5}
\begin{align}
{{b}_{1}}=&2{{\kappa }_{eff}}+{{\gamma }_{m}}>0, \\
{{b}_{2}}=&2{{\kappa }_{eff}}(\kappa _{eff}^{2}-4{{G}^{2}}+{{\Delta }^{2}}+2{{\kappa }_{eff}}{{\gamma }_{m}}) \nonumber \\
  &+{{\gamma }_{m}}(2{{\kappa }_{eff}}{{\gamma }_{m}}+\omega _{m}^{2})>0,\\
   {{b}_{3}}=&{{\omega }_{m}}[4g_{0}^{2}\Delta {{\left| {{a}_{s}} \right|}^{2}}-4g_{0}^{2}iG(a_{s}^{2}{{e}^{-i\theta }}-a_{s}^{*2}{{e}^{-i\theta }}) \nonumber\\ & +{{\omega }_{m}}(\kappa _{eff}^{2}-4{{G}^{2}}+{{\Delta }^{2}})]{{(2{{\kappa }_{eff}}+{{\gamma }_{m}})}^{2}} \nonumber\\ 
 & +[2\kappa _{eff}^{2}\omega _{m}^{2}+(\kappa _{eff}^{2}-4{{G}^{2}}+{{\Delta }^{2}}){{\gamma }_{m}}] \nonumber\\ 
   & \times [2{{\kappa }_{eff}}(\kappa _{eff}^{2}-4{{G}^{2}}+{{\Delta }^{2}})\nonumber\\
 &+(4\kappa _{eff}^{2}+\omega _{m}^{2}){{\gamma }_{m}}+2{{\kappa }_{eff}}{{\gamma }_{m}}]>0,
 \end{align}
 \begin{align}
  {{b}_{4}}=&{{\kappa }_{eff}}^{2}{{\omega }_{m}}-4{{G}^{2}}{{\omega }_{m}}+{{\Delta }^{2}}{{\omega }_{m}} \nonumber\\ 
 &-4g_{0}^{2}[{{\left| {{a}_{s}} \right|}^{2}}\Delta +iG(a_{s}^{2}{{e}^{-i\theta }}-a_{s}^{*2}{{e}^{i\theta }})]>0. 
 \end{align}
\end{subequations}
Note that the stability conditions Eq. \eqref{eq:group5} depends effective cavity decay rate ${{\kappa }_{eff}}$ with ${{\kappa }_{eff}}=\kappa ({1-r})/({1+r})$, hence on amplitude reflectivity $r$ of the BS. All the parameters are selected to satisfy these conditions in the following numerical simulation.

Applying Fourier transformation (FT) onto Eq. \eqref{eq:group4}, we could obtain the linear equations in the frequency domain. We use the FT form of $f(\omega ) = \int_{ - \infty }^{ + \infty } {f(t)} {e^{i\omega t}}dt$, ${f^\dag }(\omega ) = \int_{ - \infty }^{ + \infty } {{f^\dag }(t)} {e^{i\omega t}}dt$ and ${[{f^\dag }(\omega )]^\dag } = f( - \omega )$. By solving the equations, the position fluctuation of the movable mirror is obtained 
\begin{equation}\label{eq:16}
\delta \hat{Q}(\omega )={{A}_{1}}(\omega )\delta {{\hat{a}}_{in}}(\omega )+{{A}_{2}}(\omega )\delta \hat{a}_{in}^{\dagger }(-\omega )\text{+}{{\text{A}}_{3}}(\omega )\hat{\xi }(\omega ),
\end{equation}
where
\begin{subequations}\label{{eq:group6}}
\begin{align}
 {{A}_{1}}(\omega )=&-\frac{{{\omega }_{m}}}{d(\omega )}\Bigr[\frac{2\sqrt{2\kappa }}{r+1}t{{g}_{0}}\{[{{\kappa }_{eff}}-i(\Delta +\omega )]a_{s}^{*}\nonumber\label{eq:17a} \\ 
 & +2G{{e}^{-i\theta }}{{a}_{s}}\}\Bigr], \\
 {{A}_{2}}(\omega )=&-\frac{{{\omega }_{m}}}{d(\omega )}\Bigr[\frac{2\sqrt{2\kappa }}{r+1}t{{g}_{0}}\{[{{\kappa }_{eff}}+i(\Delta -\omega )]{{a}_{s}} \nonumber \\ 
 & +2G{{e}^{i\theta }}a_{s}^{*}\}\Bigr], \\
 {{A}_{3}}(\omega )=&-\frac{{{\omega }_{m}}}{d(\omega )}[{{({{\kappa }_{eff}}-i\omega )}^{2}}+{{\Delta }^{2}}-4{{G}^{2}}],
\end{align}
\end{subequations}
and
\begin{align}\label{eq:18}
   d(\omega )=&4{{\omega }_{m}}g_{0}^{2}[\Delta {{\left| {{a}_{s}} \right|}^{2}}+iG({{a}_{s}}^{2}{{e}^{-i\theta }}-a_{s}^{*2}{{e}^{i\theta }})]\nonumber \\ 
 & +({{\omega }^{2}}-\omega _{m}^{2}+i{{\gamma }_{m}}\omega )[{{({{\kappa }_{eff}}-i\omega )}^{2}}+{{\Delta }^{2}}-4{{G}^{2}}]. 
\end{align}
The first two terms of Eq. \eqref{eq:16} are related to input vacuum noises, while the last term is due to the thermal noise. The movable mirror is driven by both radiation pressure force and thermal force. Without opto-mechanical coupling (${{g}_{0}}\text{=}0$), Eq. \eqref{eq:16} can be simplified to $\delta \hat{Q}(\omega )=-\frac{{{\omega }_{m}}}{{{\omega }^{2}}-\omega _{m}^{2}+i{{\gamma }_{m}}\omega }\hat{\xi }(\omega )$, then the position fluctuation of movable mirror $\delta \hat{Q}(\omega )$ is only determined by the thermal Brownian noise $\hat{\xi }(\omega )$ due to the environment. We consider the symmetrized noise spectrum, and define the position spectrum of the movable mirror as 
\begin{equation}\label{eq:19}
2\pi {{S}_{Q}}(\omega )\delta (\omega +\Omega )=\frac{1}{2}\left[ \langle \delta \hat{Q}(\omega )\delta \hat{Q}(\Omega ) \rangle +\langle \delta \hat{Q}(\Omega )\delta \hat{Q}(\omega ) \rangle  \right].
\end{equation}
Take the Fourier transform of Eqs. \eqref{eq:3}, \eqref{eq:group2}, we get
\begin{subequations}\label{eq:group7}
\begin{gather}
\langle \delta {{{\hat{a}}}_{in}}(\omega )\delta \hat{a}_{in}^{\dagger }(-\Omega ) \rangle =2\pi \delta (\omega +\Omega ),\\
\langle \hat{\xi }(\omega )\hat{\xi }(\Omega ) \rangle =4\pi \frac{{{\gamma }_{m}}}{{{\omega }_{m}}}\omega \left[ 1+\coth \left( \frac{\hbar \omega }{2{{k}_{B}}T} \right) \right]\delta (\omega +\Omega ). 
\end{gather}
\end{subequations}
Inserting Eqs. \eqref{eq:16} and \eqref{eq:group7} into \eqref{eq:19}, the position spectrum of the movable mirror can be given by \cite{Huang2009a},
\begin{align}\label{eq:21}
  {{S}_{Q}}(\omega )=&\frac{1}{2}{{A}_{1}}(\omega ){{A}_{2}}(-\omega )+\frac{1}{2}{{A}_{2}}(\omega ){{A}_{1}}(-\omega ) \nonumber\\ & +2\frac{{{\gamma }_{m}}}{{{\omega }_{m}}}\omega \coth \left( \frac{\hbar \omega }{2{{k}_{B}}T} \right){{A}_{3}}(\omega ){{A}_{3}}(-\omega ),
\end{align}
and finally expressed as
\begin{align}\label{eq:22}
  {{S}_{Q}}(\omega )=&\frac{\omega _{m}^{2}}{{{\left| d(\omega ) \right|}^{2}}}\biggr\{8{{[(1-r)t]}^{2}}g_{0}^{2}\kappa [({{\kappa }_{eff}}^{2}+{{\omega }^{2}}+{{\Delta }^{2}}\nonumber\\
  &+4{{G}^{2}}){{\left| {{a}_{s}} \right|}^{2}}+2G{{e}^{i\theta }}a_{s}^{*2}({{\kappa }_{eff}}-i\Delta )\nonumber\\
 &+2G{{e}^{-i\theta }}a_{s}^{2}({{\kappa }_{eff}}+i\Delta )] \nonumber\\ 
 & +2\frac{{{\gamma }_{m}}}{{{\omega }_{m}}}\omega [{{({{\Delta }^{2}}+{{\kappa }_{eff}}^{2}-{{\omega }^{2}}-4{{G}^{2}})}^{2}}\nonumber\\
 & +4{{\kappa }_{eff}}^{2}{{\omega }^{2}}] \times \coth \left( \frac{\hbar \omega }{2{{k}_{B}}T}\right )\biggl\}.
\end{align}

The spectrum ${{S}_{Q}}(\omega )$ has two contributions. The first two terms related to ${{g}_{0}}$ is from optomechanical interaction between the movable mirror and intracavity field via radiation pressure force, while the third term is from the coupling between the movable mirror and thermal bath. Note that the position spectrum depends on three effects: the optomechanical, the OPA and the CF effects, which could be engineered by adjusting several parameters such as input optical power, OPA gain and phase, and CF strength in terms of amplitude reflectivity of the BS.

\section{Normal-mode splitting with OPA and CF}\label{normal mode splitting}
We now simulate the NMS with various parameters. In order to investigate the spectra of the movable mirror and the output field, we need to analyze the eigenvalues of $iM$ as the solution of Eq. \eqref{eq:11} in the frequency domain or to study the complex zeroes of the function $d(\omega )$ in Eq. \eqref{eq:18}. We use the latter to numerically simulate the NMS including its frequency separation and linewidth, which are supposed to depend on the effective cavity detuning $\Delta $, parametric gain $G$ and phase, the steady-state amplitude of the cavity field, which is related to input power and the BS amplitude reflectivity $r$.

We use parameters similar to those in an experiment about observing the NMS of the fluctuation spectra \cite{Grblacher2009}: $\lambda={2\pi c}/{\omega }_{l}=1064\ \mathrm {nm}$, $L=25\ \mathrm {mm}$, $m=145\ \mathrm{ng}$, $\kappa =2\pi\times215\times {{10}^{3}}\ \mathrm {Hz}$, ${{\omega }_{m}}=2\pi \times 947\times {{10}^{3}}\ \mathrm {Hz}$, $T\text{=}300\ \mathrm {mK}$ and the mechanical quality factor ${Q}'={{{\omega }_{m}}}/{{{\gamma }_{m}}}=6700$. In the high-temperature limit ${{k}_{B}}T\gg \hbar {{\omega }_{m}}$, we can take an approximation $\coth \left( {\hbar {{\omega }_{m}}}/{2{{k}_{B}}T} \right)\approx {2{{k}_{B}}T}/{\hbar {{\omega }_{m}}}$. We consider the degenerate case $\Delta=\sqrt{\omega_{m}{ }^{2}+4 G^{2}}$ which is the most efficient for the coupling between two normal modes \cite{Huang2009}.
\begin{figure}[htbp]
    \centering
    \includegraphics[width=\linewidth]{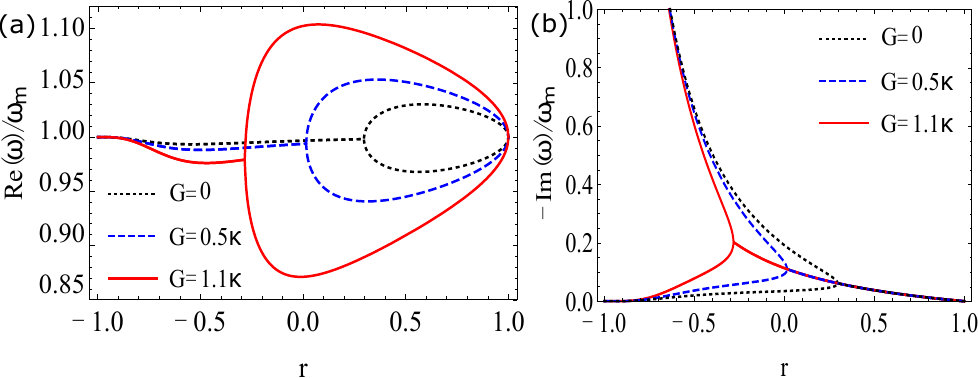} 
    \caption{(a) and (b) are the real parts of the solutions of $d(\omega )$ in the domain $\operatorname{Re}(\omega )>0$ and the imaginary parts of the solutions of $d(\omega )$, respectively, as a function of amplitude reflectivity. $G=0$ (dotted curve), $G=0.5\kappa $ (dashed curve), $G=1.1\kappa $ (solid curve). Parameters: $P_{\text {in }}=2\ \mathrm{mW}$, $\theta=-\pi/4$.} 
    \label{FIG.2}
\end{figure}

Setting $d(\omega)=0$ in Eq. \eqref{eq:18} we could get complex  solutions of $\omega$, whose real (imaginary) parts stand for the frequencies (linewidths) of resonant modes (normal modes). In Fig. \ref{FIG.2}, we plot the real parts and imaginary parts for the parametric gain $G=0$, $G=0.5\kappa $ and $G=1.1\kappa$ versus the amplitude reflectivity $r$ of the BS with OPA phase of $\theta=-\pi/4$. Without OPA ($G=0$), when $r<0.3$, $\operatorname{Re}(\omega )$ has two identical values, i.e., no NMS occurs, while $\operatorname{Im}(\omega )$ has two unequal roots, showing the linewidth splitting appears \cite{Gupta1995}. With $r>0.3$ and $r$ increasing, NMS occurs and the distance between two normal modes increases at the begining and then decreases and becomes zero at $r=1$, which is reasonable as there is no light in the cavity, meanwhile, the linewidth splitting starts disappearing with the same value of $r$ to that of NMS occuring and then merged linewidth decreases with increasing $r$. Adding OPA($G\neq 0 $), particularly, $G=0.5\kappa$, NMS occurs at $r=0.015$. With the gain increasing, say $G=1.1\kappa $, the NMS begin to appear at $r=-0.3$ and the distance of two normal modes increased. To sum up, (1) OPA enhances the NMS in general, because photon number in the cavity increases with increasing gain, leading to a stronger optomechanical coupling between the movable mirror and the cavity field. (2) CF can adjust the NMS with large scale, by changing the strength (absolute value of $r$) of CF or the feedback direction (the plus or minus of $r$), in this way changing the optomechanical coupling strength by using both OPA and CF, adding flexibility of control.

\begin{figure}[htbp]
    \centering
    \includegraphics[width=\linewidth]{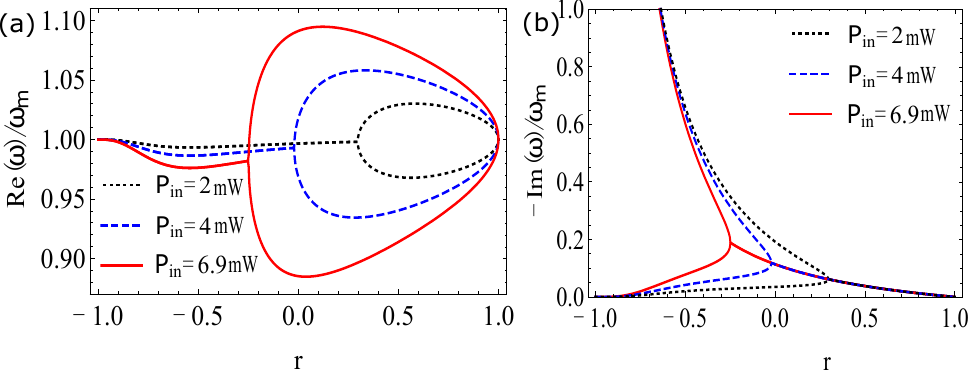} 
    \caption{(a) and (b) are are the real parts of the solutions of $d(\omega )$ in the domain $\operatorname{Re}(\omega )>0$ and the imaginary parts of the solutions of $d(\omega )$, respectively, as a function of amplitude reflectivity with $P_{\text {in }}=2\ \mathrm {mW}$ (dotted curve), $P_{\text {in }}=4\ \mathrm {mW}$ (dashed curve), $P_{\text {in }}=6.9\ \mathrm {mW}$ (solid curve). Parameters: $G=0, \ \theta=-\pi/4$.}
    \label{FIG.3}
\end{figure}

With different input laser powers $P_{\text{in}}$ of $2\ \mathrm{mW}$, $4\ \mathrm{mW}$ and $6.9\ \mathrm{mW}$, the real and imaginary parts of $\operatorname{Re}(\omega )$ versus BS amplitude reflectivity $r$ are plotted in Fig. \ref{FIG.3}. The NMS start to appear when $r$ is $0.3$, $-0.02$ and $-0.25$, respectively, showing increased optical power enhances the NMS: increasing the separation of two normal modes and the NMS scale across $r$, and showing very similar behavior to that for different OPA gains in Fig. \ref{FIG.2}. The reason behind this is that increasing input laser power leads to a larger light-enhanced optomechanical coupling strength, equivalent to increasing OPA gain.

\begin{figure}[htbp]
    \centering
    \includegraphics[width=\linewidth]{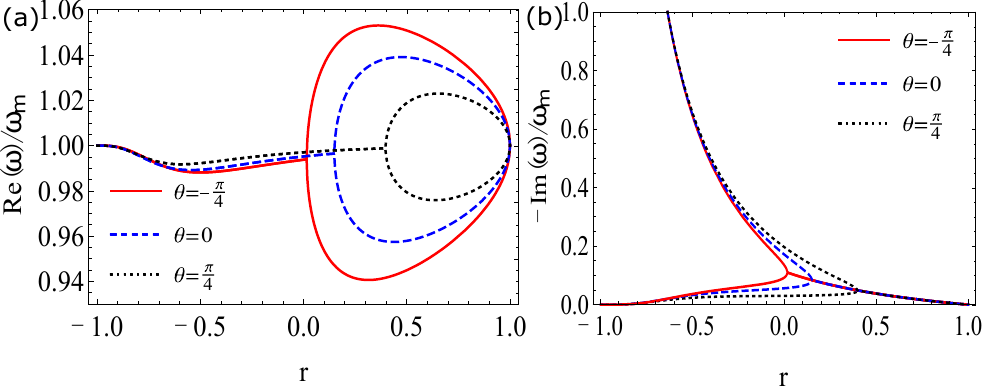}
    \caption{(a) and (b) are are the real parts of the solutions of $d(\omega )$ in the domain $\operatorname{Re}(\omega )>0$ and the imaginary parts of the solutions of $d(\omega )$, respectively, as a function of amplitude reflectivity $r$ of the beam splitter for different parametric phase. $\theta=\pi/4$ (dotted curve), $\theta=0$ (dashed curve), $\theta=-\pi/4$ (solid curve). Parameters: $G=0.5\kappa $, ${{P}_{\text{in}}}\text{=}2$ {mW}.}
    \label{FIG.4}
\end{figure}

We also study the effect of OPA phase $\theta$ on the NMS, shown in Fig. \ref{FIG.4}. It shows again a very similar behavior to that of OPA gain and input laser power. Insterstingly, the better NMS occurs at $\theta=-\pi/4$ (red solid), not in a general amplifying state at $\theta=0$ (blue dashed). This could be explained by Eq. \eqref{eq:9c}, the steady-state intracavity optical amplitude. The effective cavity detuning ($\Delta\neq 0$) changes the resonant condition of intracavity round-trip phase, thus changing the OPA amplification (in-phase) and deamplification (out-of-phase) condition. The intracavity round-trip phase with $G=0.5\kappa$ is expressed as $-\text{arctan(}{\Delta }/{{{\kappa }_{eff}}})$, which approaches $-\pi/4$, hence showing more obvious NMS than that of $\theta=0$ and $\theta=\pi/4$.

\section{Noise spectra of the position of movable mirror and output field}\label{spectrum}

The NMS of the position of movable mirror could be observed by the spectra of the output field of optomechanical system, which is obtained by Fourier transforming the fluctuation equation of motion of the cavity field, i.e., Eq. \eqref{eq:group4} and then solving the corresponding linear equations. Then the intracavity field fluctuation is given by
 \begin{equation}\label{eq:25}
   \delta \hat{a}(\omega )={{B}_{1}}(\omega )\delta {{\hat{a}}_{in}}(\omega )+{{B}_{2}}(\omega )\delta \hat{a}_{in}^{\dagger }(-\omega )\text{+}{{\text{B}}_{3}}(\omega )\hat{\xi }(\omega).
 \end{equation}
Using the input-output relationship, Eq. \eqref{eq:5}, the relationship of fluctuation terms between intra-cavity field and outfield is $\delta {{\hat{{a}'}}_{out}}=\frac{\sqrt{2\kappa }}{1+r}\delta \hat{a}-\frac{t}{1+r}\delta {{\hat{a}}_{in}}$, the fluctuations of the output field can be written as,
\begin{align}\label{eq:26}
  \delta {{{\hat{{a}'}}}_{out}}(\omega )={{C}_{1}}(\omega )\delta {{{\hat{a}}}_{in}}(\omega )+{{C}_{2}}(\omega )\delta \hat{a}_{in}^{\dagger }(-\omega )\text{+}{{\text{C}}_{3}}(\omega )\hat{\xi }(\omega ),
\end{align}
and the final output field of BS through the relation ${{\hat{a}}_{out}}\text{=}t{{\hat{{a}'}}_{out}}-r{{\hat{a}}_{in}}$ is given by
\begin{equation}\label{eq:27}
\delta {{\hat{a}}_{out}}(\omega )={{D}_{1}}(\omega )\delta {{\hat{a}}_{in}}(\omega )+{{D}_{2}}(\omega )\delta \hat{a}_{in}^{\dagger }(-\omega )\text{+}{{\text{D}}_{3}}(\omega )\hat{\xi }(\omega ),
\end{equation}
with ${{C}_{1}}(\omega )\text{=}{\sqrt{2\kappa }}/({1+r}){{B}_{1}}(\omega)-{t}/(1+r),$
${{C}_{2}}(\omega )=\sqrt{2\kappa }/(1+r){{B}_{2}}(\omega )$, and ${{C}_{3}}(\omega )\text{=}{\sqrt{2\kappa }}/({1+r}){{B}_{3}}(\omega )$. Likely, ${{D}_{1}}(\omega )\text{=}t{{C}_{1}}(\omega)-r$, ${{D}_{2}}(\omega )\text{=}t{{C}_{2}}(\omega )$ and ${{D}_{3}}(\omega )\text{=}t{{C}_{3}}(\omega )$,
where
\begin{subequations}\label{eq:group8}
\begin{align}
 {{D}_{1}}(\omega )=&\frac{2\kappa {{t}^{2}}}{{{(1+r)}^{2}}{{({{\kappa }_{eff}}-i\omega )}^{2}}+{{\Delta }^{2}}-4{{G}^{2}}}\nonumber \\ 
 & \times \Bigr[-\frac{2\omega _{m}g_{0}^{2}}{d(\omega )}i\{[{{\kappa }_{eff}}-i(\Delta +\omega )]{{a}_{s}}-2G{{e}^{i\theta }}a_{s}^{*}\} \nonumber\\ 
 & \{[{{\kappa }_{eff}}-i(\Delta +\omega )]a_{s}^{*}+2G{{e}^{-i\theta }}{{a}_{s}}\}\nonumber \\ 
 & +{{\kappa }_{eff}}-i(\Delta +\omega )\Bigm]-\frac{{{t}^{2}}}{1+r}-r,\\
 {{D}_{2}}(\omega )=&\frac{2\kappa {{t}^{2}}}{{{(1+r)}^{2}}{{({{\kappa }_{eff}}-i\omega )}^{2}}+{{\Delta }^{2}}-4{{G}^{2}}}\nonumber\\
  & \times \Bigr[-\frac{2{{\omega }_{\text{m}}}g_{0}^{2}}{d(\omega )}i\{[{{\kappa }_{eff}}-i(\Delta +\omega )]{{a}_{s}}-2G{{e}^{i\theta }}a_{s}^{*}\} \nonumber\\ 
 & \{[{{\kappa }_{eff}}-i(\omega -\Delta )]{{a}_{s}}+2G{{e}^{i\theta }}a_{s}^{*}\}+2G{{e}^{i\theta }}\Bigm],\\
{{D}_{3}}(\omega )=&-\frac{\sqrt{2\kappa }\omega _{m}{{g}_{0}}t}{(1+r)d(\omega )}i\{[{{\kappa }_{eff}}-i(\omega +\Delta )]{{a}_{s}}\nonumber\\
&-2G{{e}^{i\theta }}a_{s}^{*}\}.
\end{align}
\end{subequations}
The amplitude and phase quadratures of $X$ and $Y$ are two commonly used observables. With the fluctuations of $\delta {{x}_{out}}(\omega )=\delta {{\hat{a}}_{out}}(\omega )+\delta \hat{a}_{out}^{\dagger }(\omega )$ and $\delta {{y}_{out}}(\omega )=i[\delta \hat{a}_{out}^{\dagger }(\omega )-\delta {{\hat{a}}_{out}}(\omega )]$, the spectra of the output field are difined as
\begin{subequations}\label{eq:group9}
\begin{align}\label{eq:29}
2\pi {{S}_{aout}}(\omega )\delta (\omega +\Omega )=\langle \delta \hat{a}_{out}^{\dagger }(-\Omega )\delta {{{\hat{a}}}_{out}}(\omega ) \rangle,\\
2\pi {{S}_{xout}}(\omega )\delta (\omega +\Omega )=\left\langle \delta {{x}_{out}}(\Omega )\delta {{x}_{out}}(\omega ) \right\rangle,\\
2\pi {{S}_{yout}}(\omega )\delta (\omega +\Omega )=\left\langle \delta {{y}_{out}}(\Omega )\delta {{y}_{out}}(\omega ) \right\rangle. 
\end{align}
\end{subequations}
Combining Eq. \eqref{eq:27} and the correlation functions of the noise operators in frequency domain, we get the final expressions of spectra
\begin{subequations}\label{eq:group10}
\begin{align}\label{eq:30}
   {{S}_{aout}}(\omega )=& D_{2}^{*}(\omega ){{D}_{2}}(\omega )+D_{3}^{*}(\omega ){{D}_{3}}(\omega ) \nonumber\\ 
 & \times 2\frac{{{\gamma }_{m}}}{{{\omega }_{m}}}\omega \coth \left[ -1+\frac{\hbar \omega }{2{{k}_{B}}T} \right],\\  
   {{S}_{xout}}(\omega )=&[{{D}_{1}}(-\omega )+D_{2}^{*}(\omega )][{{D}_{2}}(\omega )+D_{1}^{*}(-\omega )] \nonumber\\ 
 & +[{{D}_{3}}(-\omega )+D_{3}^{*}(\omega )][{{D}_{3}}(\omega )+D_{3}^{*}(-\omega )]\nonumber \\ 
 & \times 2\frac{{{\gamma }_{m}}}{{{\omega }_{m}}}\omega \coth \left[ -1+\frac{\hbar \omega }{2{{k}_{B}}T} \right], \\
   {{S}_{yout}}(\omega )=&-[D_{2}^{*}(\omega )-{{D}_{1}}(-\omega )][D_{1}^{*}(-\omega )-{{D}_{2}}(\omega )] \nonumber\\ 
 & -[D_{3}^{*}(\omega )-{{D}_{3}}(-\omega )][D_{3}^{*}(-\omega )-{{D}_{3}}(\omega )] \nonumber\\ 
 & \times 2\frac{{{\gamma }_{m}}}{{{\omega }_{m}}}\omega \coth \left[ -1+\frac{\hbar \omega }{2{{k}_{B}}T} \right]. 
\end{align}
\end{subequations}
For any single spectrum above, there are two contributions, one from the input vacuum noise and the other from the thermal noise of the coupling between the movable mirror and the thermal bath. As shown in Ref. \cite{Huang2009}, these three terms exhibit similar spectra behaviors, thus only $S_{acout}$ is plotted in the following text.

\begin{figure}[htbp]
    \centering
    \includegraphics[width=\linewidth]{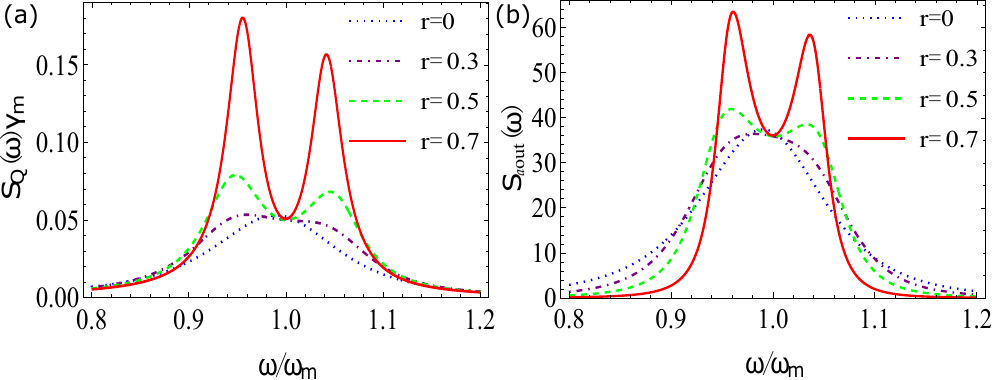}
    \caption{(Color online) (a) the position spectra ${{S}_{Q}}(\omega )\times {{\gamma }_{m}}$ of the movable mirror and (b) the spectra ${{S}_{aout}}(\omega )$ of the output field, versus the normalized frequency $\omega /{{\omega }_{m}}$ for different BS amplitude reflectivity $r=0$ (dotted curve),  $r=0.3$ (dot-dashed),  $r=0.5$ (dashed curve) and $r=0.7$ (solid curve), with the same other parameters: $G=0.5\kappa $, ${{P}_{\text{in}}}\text{=}2\ \mathrm {mW},  \theta=-\pi/4$.}
    \label{FIG.5}
\end{figure}
According to Eqs. \eqref{eq:22} and \eqref{eq:30}, the position spectra of the movable mirror and the output field spectra are shown in Fig. \ref{FIG.5}, with different CF strength and the same OPA gain $G=0.5\kappa $. Fig. \ref{FIG.5} (a) and (b) respectively show the spectra ${{S}_{Q}}(\omega )$ and ${{S}_{aout}}(\omega )$ versus the normalized frequency $\omega /{{\omega }_{m}}$ for different amplitude reflectivity $r$ of the BS in the presence of OPA. As expected, the NMS phenomena are not obvious without coherent feedback, as amplitude reflectivity of BS increases, the interval between two peaks first becomes lager and then becomes smaller, and the peaks become higher and narrower.

\begin{figure}[htbp]
    \centering
    \includegraphics[width=\linewidth]{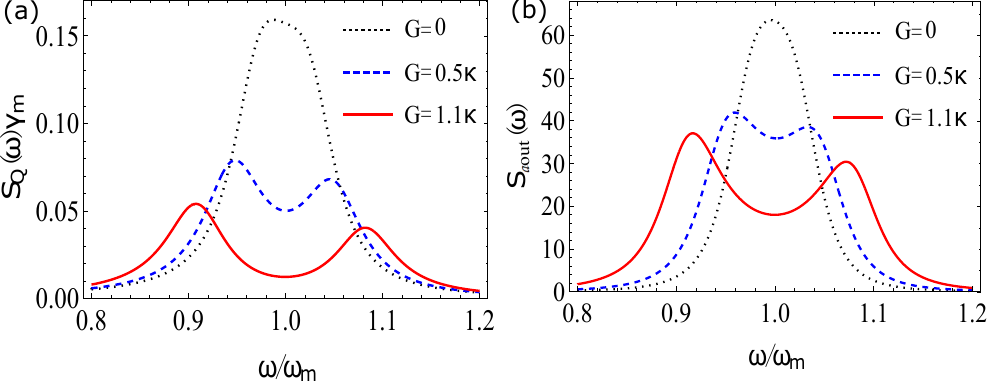}
    \caption{(Color online) (a) the position spectra ${{S}_{Q}}(\omega )\times {{\gamma }_{m}}$ of the movable mirror and (b) the spectra ${{S}_{aout}}(\omega )$ of the output field, versus the normalized frequency $\omega /{{\omega }_{m}}$ for different OPA gain $G=0$ (dotted curve), $G=0.5\kappa $ (dashed curve) and $G=1.1\kappa $ (solid curve) with the same other parameters: $\theta=-\pi/4$, $r=0.5$, ${{P}_{\text{in}}}\text{=}2\ \mathrm {mW}$.}
    \label{FIG.6}
\end{figure}

Fig. \ref{FIG.6} shows the spectra ${{S}_{Q}}(\omega )$ and ${{S}_{aout}}(\omega )$ with different OPA gains and the same CF strength ($r=0.5$). The peak separation becomes lager with increasing gain, hence the NMS becomes more obvious. Figs. \ref{FIG.5} and \ref{FIG.6} demonstrate that a combination of OPA and CF makes it easier to enter the strongly coupling domain and enhances the NMS, and adding controlling flexibility compared to that with OPA or CF alone.

\section{Conclusion}\label{conclusion}
In conclusion, we have theoretically studied the normal-mode splitting in an optomechanical system with OPA and CF. We analyzed the NMS with different input power, CF strength, OPA gain and phase. Compared with the previous scheme with OPA or CF alone, the NMS can be optimized and enhanced by adjusting the above parameters with more flexibility. This scheme may be extended to a variety of macroscopic quantum systems such as cavity electromechanics \cite{Blais2021} and optomagnonics \cite{Li2021}.

\begin{acknowledgments}
We thank Heng Shen and Zhihui Yan for discussions on CF and NMS, and Pengfei Zhang on vacuum Rabi splitting. This work is supported by the Research Project Supported by Shanxi Scholarship Council of China (2021-005), the Ministry of Science and Technology of the People’s Republic of China (MOST) (Grant No. 2021YFC2201802), the National Natural Science Foundation of China (Grants Nos. 91536222, 12074233) and Shanxi 1331 Project.
\end{acknowledgments}

\appendix
\counterwithin{figure}{section}
\section{Derivation of Input-output Relation for Coherent Feedback Cavity}\label{appendix}

\begin{figure}[htbp]
    \centering
    \includegraphics[width=7cm]{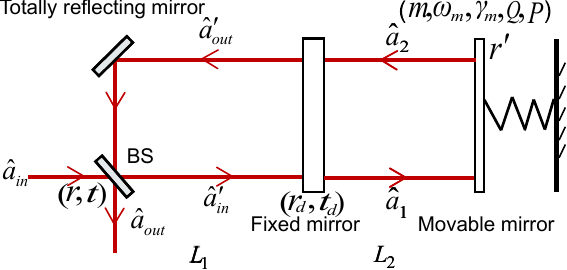} 
    \caption{Schematic of an optical cavity with CF}
    \label{FIG.appendix}
\end{figure}

The input-output relation of cavity after adding feedback is derived as follows \cite{2008Dispersive}.  The basic scheme is shown in Fig. \ref{FIG.appendix}, where OPA and optomechanical coupling could be ignored in this derivation.

The equations of field relations are given by
\begin{subequations}
\begin{align}
  & {{{\hat{{a}'}}}_{in}}=it{{{\hat{a}}}_{in}}+r{{{\hat{{a}'}}}_{out}}{{e}^{ik{{L}_{1}}}}, \\ 
 & {{{\hat{{a}'}}}_{out}}={{r}_{d}}{{{\hat{{a}'}}}_{in}}{{e}^{ik{{L}_{1}}}}+i{{t}_{d}}{{{\hat{a}}}_{2}}{{e}^{ik{{L}_{2}}}}, \\ 
 & {{{\hat{a}}}_{1}}=i{{t}_{d}}{{{\hat{{a}'}}}_{in}}{{e}^{ik{{L}_{1}}}}+{{r}_{d}}{{{\hat{a}}}_{2}}{{e}^{ik{{L}_{2}}}}, \\ 
 & {{{\hat{a}}}_{2}}={r}'{{{\hat{a}}}_{1}}{{e}^{ik{{L}_{2}}}}.
\end{align}
\end{subequations}
Here ${{L}_{1}}$ and ${{L}_{2}}$ are the round-trip lengths of the left- and right-hand halves of the cavity, $k$ is the wavenumber of laser light, $r$ (${{r}_{d}}$) and $t$ (${{t}_{d}}$) are the amplitude reflectivity and transmissivity of the BS (fixed mirror). The movable mirror is regarded as a totally reflecting mirror (${r}'=1$). The following assumptions are made: the transmitted and refleted beams of the BS are in-phase or out-of-phase interfered ($-1\leqslant r\leqslant 1$ and $r\in Reals$)  by making (${{e}^{ik{{L}_{1}}}}\text{=}i$). And the cavity field resonates with the input laser (${{e}^{ik{{L}_{2}}}}\text{=}-1$). Then the equations is simplified:
\begin{subequations}\label{eq:group11}
\begin{align}\label{eq.A2a} 
  & {{{\hat{{a}'}}}_{in}}=it{{{\hat{a}}}_{in}}+ir{{{\hat{{a}'}}}_{out}}, \\\label{eq.A2b} 
 & {{{\hat{{a}'}}}_{out}}=i{{r}_{d}}{{{\hat{{a}'}}}_{in}}-i{{t}_{d}}{{{\hat{a}}}_{2}}, \\ \label{eq.A2c}
 & {{{\hat{a}}}_{1}}={{-t}_{d}}{{{\hat{{a}'}}}_{in}}-{{r}_{d}}{{{\hat{a}}}_{2}}, \\ \label{eq.A2d}
 & {{{\hat{a}}}_{2}}={{{-\hat{a}}}_{1}}.
\end{align}
\end{subequations}
Combining Eq. \eqref{eq.A2a} with Eq. \eqref{eq.A2b}, the input-output relation is given by 
\begin{equation}\label{eq.A3}
{{\hat{{a}'}}_{out}}=\frac{-i{{t}_{d}}}{1+r{{r}_{d}}}\hat{a}-\frac{{{r}_{d}}}{1+r{{r}_{d}}}t{{\hat{a}}_{in}}.
\end{equation}
Substituting Eq. \eqref{eq:group11} into Eq. \eqref{eq.A3}, the intra-cavity field is written as 
\begin{equation}\label{A4}
 \hat{a}\text{=}\frac{i{{t}_{d}}}{1+r{{r}_{d}}-r-{{r}_{d}}}t{{\hat{a}}_{in}}, 
\end{equation}
With small transmission loss $\gamma $, the following relation is satisfied,
\begin{subequations}\label{eq:group12}
\begin{align}
 {{r}_{d}}\approx 1-\gamma,\\    
 {{t}_{d}}=\sqrt{2\gamma },  
\end{align}
\end{subequations}
where $\gamma =\kappa \tau $, $\tau $  is the light round-trip time of the right-hand cavity. The cavity field is expressed as 
\begin{equation}
\hat{a}=\frac{-i\sqrt{2\kappa}}{(1-r)\kappa}t\left(\frac{{\hat{a}}_{in}}{\sqrt{\tau}}\right),
\end{equation}
where $\left(\frac{{\hat{a}}_{in}}{\sqrt{\tau}}\right)\equiv {\hat{a}}^{norm}_{in}$ is in fact the time-normalized input field amplitude.
which is in agreement with the expression of the steady-state intracavity field of Eq. \eqref{eq:9c} in the case of resonance ($\Delta=0$) and without OPA( $G=0$) in the main text
\begin{equation}
{{a}_{s}}\text{=}\frac{\sqrt{2\kappa }}{(1\text{-}r)\kappa }t{{{a}}_{in}}.
\end{equation}


\bibliography{NMS}

\end{document}